\documentclass[pre,showpacs,amsmath,twocolumn]{revtex4}  %% REVTeX 4.0

\usepackage{graphicx}% Include figure files
\usepackage{amssymb}
\usepackage{bm}

\begin{document}
\title{Macroscopic dynamics of biological cells interacting
via chemotaxis and direct contact}

\author{Pavel M. Lushnikov$^{1*}$,  Nan Chen$^2$ and Mark
 Alber$^{2}$
}

\affiliation{$^1$Department of Mathematics and Statistics,
University of New Mexico, Albuquerque, NM 87131, USA   \\
%  $^2$ Landau Institute for Theoretical Physics, Kosygin St. 2,
%  Moscow, 119334, Russia
%\\
$^2$Department of Mathematics, University of Notre
Dame, Notre Dame,
46656, USA
}

\email{plushnik@math.unm.edu}

\date{%Printed
\today
%July 20, 2008
}

\begin{abstract}
A connection is established between discrete stochastic model
describing microscopic motion of fluctuating  cells, and macroscopic
equations describing dynamics of cellular density. Cells move
towards  chemical gradient (process called chemotaxis)
with their shapes randomly fluctuating. Nonlinear  diffusion equation is derived from microscopic
dynamics in dimensions one and two using excluded volume approach.
Nonlinear diffusion coefficient depends on cellular volume fraction
and it is demonstrated to prevent collapse of cellular density. A
very good agreement is shown between Monte Carlo simulations of the
microscopic Cellular Potts Model and numerical solutions of the
macroscopic equations for relatively large cellular volume
fractions. Combination of  microscopic and  macroscopic models were
used to simulate growth of structures similar to early vascular
networks.
\end{abstract}


\maketitle

* author for correspondence: Pavel Lushnikov

\section{Introduction}
So far most models used in biology have been developed at specific
scales. Establishing a connection between discrete stochastic
microscopic description and continuous deterministic macroscopic
description of the same biological phenomenon would allow one to
switch when needed from one scale to another, considering events at
individual (microscopic) cell level such as cell-cell interaction or
cell division to events involving thousands of cells such as organ
formation and development. Due to the fast calculation speed
possible with the continuous model, one can quickly test wide
parameter ranges and determine satiability conditions and then use
this information for running Monte Carlo simulations of the
stochastic discrete dynamical systems. Also, continuous models
provide very good approximation for systems containing a
biologically realistic (i.e., large) number of cells, for which
numerical simulations of stochastic trajectories can be prohibitive.

Most continuous biological models have been  postulated either by
requiring certain biologically relevant features from the solutions
or making it easier to analyze behavior of solutions using certain
mathematical techniques. In particular, system of nonlinear partial
differential equations model with chemotactic term was used in
\cite{gamba, Serini} to simulate the de novo blood vessel formation
from the mesoderm. The rational for the model was provided by the
experimental observations \cite{Car} demonstrating that chemotaxis
played an important role in guiding cells during early vascular
network formation. Discrete models have been also applied to
simulating vasculogenesis and angeogenesis \cite{Merks,rupp,szabo}.

In this paper we derive continuous macroscopic limits of the 1D and
2D microscopic cell-based models with extended cell representations,
in the form of nonlinear diffusion equations coupled with chemotaxis
equation. We demonstrate that combination of the discrete model and
derived continuous model can be used for simulating biological
phenomena in which a nonconfluent population of cells interact
directly and via diffusible factors, forming an open network
structures in a way similar to formation of networks during
vasculogenesis \cite{gamba, Serini} and pattern formation in limb
cell cultures \cite{Wu}.

Continuous limits of microscopic models of biological systems based
on point-wise based cell representation were extensively studied
over the last 30 years. The classical Keller-Segel PDE model has
been derived in \cite{KS} from a discrete model with point-wise
cells undergoing random walk in chemotactic field and then studied
in \cite{Alt1980,Othmer,StevensSIAM2000,NewmanGrima2004}. Cells in
this model secret diffusing chemical at constant rate and detect
local concentration $c$ of this chemical due to process called
chemotaxis. The chemical is called an attractant or repellent
depending whether cell moves towards  chemical
gradient or in opposite direction. Aggregation occurs if
attraction exceeds diffusion of cells. For point-wise cells
aggregation results in infinite cellular density corresponding to
the solution of the macroscopic Keller-Segel equation becoming
infinite in finite time (also called blow up in finite time or
collapse of the solution) \cite{HerreroVelazquez1996,BrennerConstantinKadanoff1999}.

There have been few attempts to derive macroscopic limits of
microscopic models which treat cells as extended objects consisting
of several points. In \cite{turner2004} the diffusion coefficient
for a collection of noninteracting randomly moving cells was derived
from a one-dimensional Cellular Potts Model (CPM). A microscopic
limit of subcellular elements model \cite{NewmanMathBioEng2005} was
derived in the form of an advection-diffusion  PDE for cellular
density. In our previous papers \cite{alber1,
alber2,AlberChenLushnikovNewmanPRL2007} we studied the continuous
limit  of the CPM describing individual cell motion in a medium and
in the presence of an external field with contact cell-cell
interactions in mean-field approximation. However, mean-field
approximation does not allow one to consider high density of cells
when cellular volume fraction (fraction of volume occupied by cells)
$\phi$ is of the order one.

\begin{figure}
\begin{center}
\includegraphics [width=60mm]{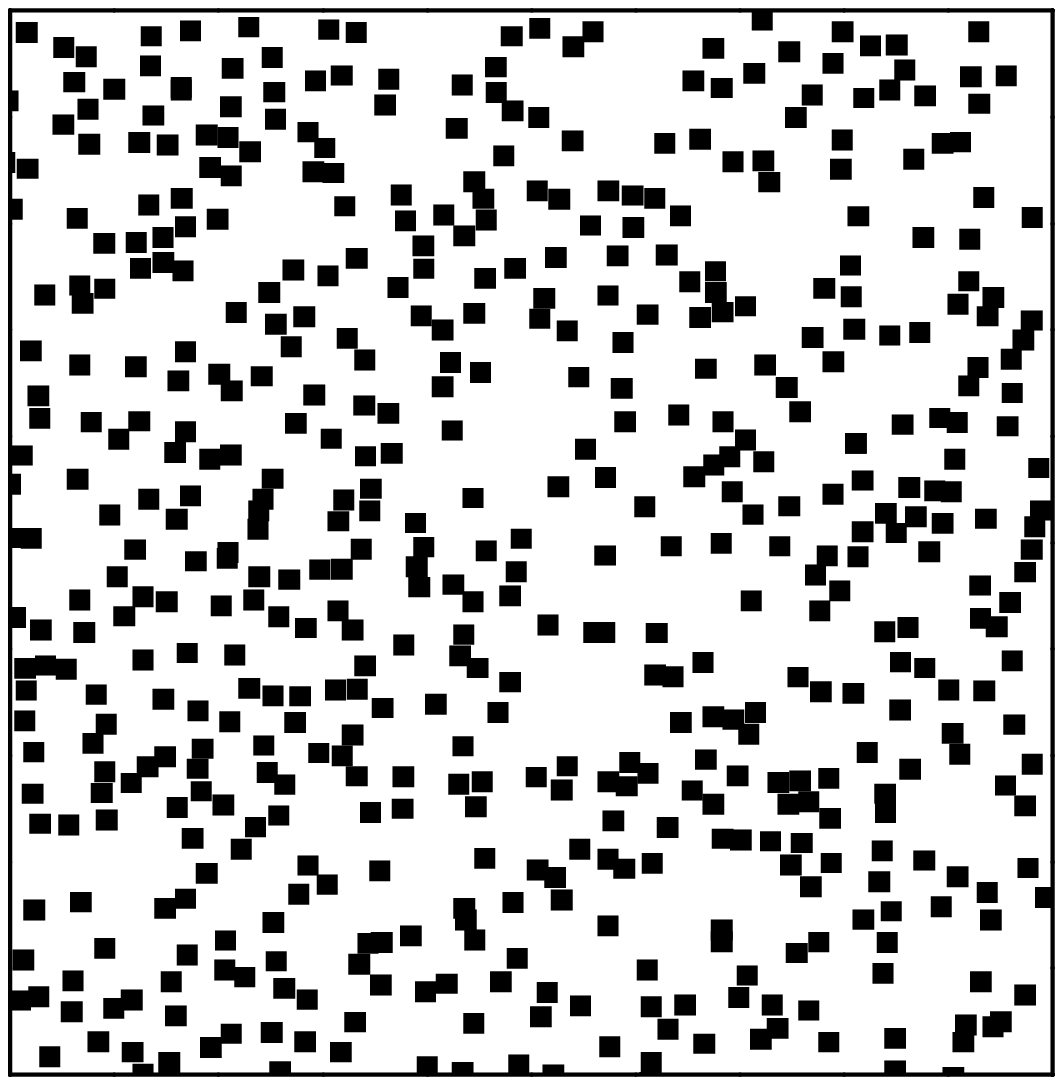}
\caption{Representation of cells in the 2D CPM in the form of
fluctuating rectangles. 539 cells (15\% volume fraction) with
$L_x^{(0)}=L_{y}^{(0)})=1.666667$ are
shown.}\label{figmultiparticles}
\end{center}
\end{figure}

In this paper we go beyond mean-field approximation. Namely, we take
into account finite size of cells in the CPM, use exclusion volume
principle (meaning that two cells can not occupy the same volume)
and derive the following macroscopic nonlinear diffusion equation
for evolution of cellular density $p({\bf r},t)$ in one (1D)
\begin{eqnarray}\label{pottscontinuous4rectangles1Dphi}
\partial _tp=D_2\nabla_{\bf r} \cdot \Big [ \frac{1+\phi^2}{(1-
\phi)^2}\nabla_{\bf r} p\Big ]-\chi_0\nabla_{\bf r} \cdot \big [p\, \nabla_{\bf
r}c({\bf r},t)\big ],
\end{eqnarray}
and two dimensions (2D):
\begin{eqnarray}\label{pottscontinuousexcluded2Dphi}
\partial _tp=D_2\nabla_{\bf r} \cdot \Big [ \frac{1+\phi}{1-
\phi+\phi\log(\phi)}\nabla_{\bf r} p\Big ]
\nonumber\\
-\chi_0\nabla_{\bf r} \cdot \big [p\, \nabla_{\bf
r}c({\bf r},t)\big ],
\end{eqnarray}
which do not have blow up in finite time. These nonlinear diffusion
equations are coupled with the equation for evolution of chemical
field $c({\bf r},t):$
\begin{eqnarray}\label{ceqmacroscopic1}
\partial_t c({\bf r},t)=D_c\nabla_{\bf r}^2   c+  ap-\gamma c.
\end{eqnarray}
Here $\phi$ is a volume fraction (fraction of volume occupied by
cells). In 1D case cells have a form of fluctuating rods and
$\phi=L_{x}^{(0)}p({\bf r},t)$, where $L_{x}^{(0)}$ is an average
length of cells. In 2D case we assume that cells are fluctuating
rectangles and $\phi=L_{x}^{(0)}L_{y}^{(0)}p({\bf r},t)$, where
$(L_{x}^{(0)},L_{y}^{(0)})$ are average length and width of a cell.
Here $p({\bf r},t)$ is the density of cells normalized by the total
number of cells $N:$ $\int p({\bf r},t)d{\bf r}=N$ and ${\bf
r}$ is a vector of spatial coordinates in 1D or 2D. $D_2$ is the
diffusion coefficient for a motion of an isolated cell, $\chi_0$
defines strength of chemotactic interactions, $D_c$ is the diffusion
coefficient of a chemical, $a$ is the production rate of a chemical
and $\gamma$ is the decay rate of a chemical. Typical microscopic
picture of distribution of individual cells is shown in Figure
\ref{figmultiparticles} for 2D case. Solutions of Eqs.
(\ref{pottscontinuous4rectangles1Dphi}),
(\ref{pottscontinuousexcluded2Dphi}) and (\ref{ceqmacroscopic1})
describe coarse-grained macroscopic cellular dynamics.
Below we show very good agreement even for relatively large
densities  $\phi\simeq0.3.$, between solutions of Eqs.
(\ref{pottscontinuous4rectangles1Dphi}),
(\ref{pottscontinuousexcluded2Dphi}), (\ref{ceqmacroscopic1}) and ensemble average of
stochastic trajectories of the microscopic CPM.

In Section II we introduce general microscopic equations describing
motion of cells based on random fluctuations of their shapes and
their interactions. We assume that each cell has a rectangular shape
and consider stochastic differential equations for motion of cells
as well as Smoluchowski equation (see e.g. \cite{Gardiner2004}) for
multi-cellular probability density function. In Section III we
consider a particular case of microscopic cellular dynamics
represented by the CPM without excluded volume interactions.
Coefficients of Smoluchowski Eq. are derived from the CPM and
stochastic dynamics of shapes and positions of cells are reduced to
solutions of the closed equation describing positions of cells. In
Section IV we consider cell motion and cell-cell interactions with
collisions resolved through the jump processes resulting in
equations (\ref{pottscontinuous4rectangles1Dphi}),
(\ref{pottscontinuousexcluded2Dphi}) and (\ref{ceqmacroscopic1}). In
Section V numerics for the continuous macroscopic equations is
compared with Monte Carlo simulations of the microscopic CPM. In
Section VI we summarize main results and discuss future directions.

\section{Microscopic  motion of cells}
Motion of many eukaryotic cells and bacteria is accompanied by the
random fluctuations of their shapes \cite{DD,CM,AM} resulting in a
random diffusion of center of mass of an isolated cell. Coefficient
of such diffusion can be measured experimentally (see e.g.
\cite{glazier2000}). Fluctuations of cellular membrane in the
presence of a chemical field are more likely in the direction of
chemical gradient (for chemoattractant) of in opposite direction
(for chemorepellent). Cells can also interact through direct contact
which includes cell-cell adhesion and can be modeled using excluded
volume principle. Cellular environment is highly viscous and inertia
of cell can be ignored.

In this paper we  assume that each cell has a fluctuating
rectangular shape
\begin{figure}
\begin{center}
\includegraphics [width=60mm]{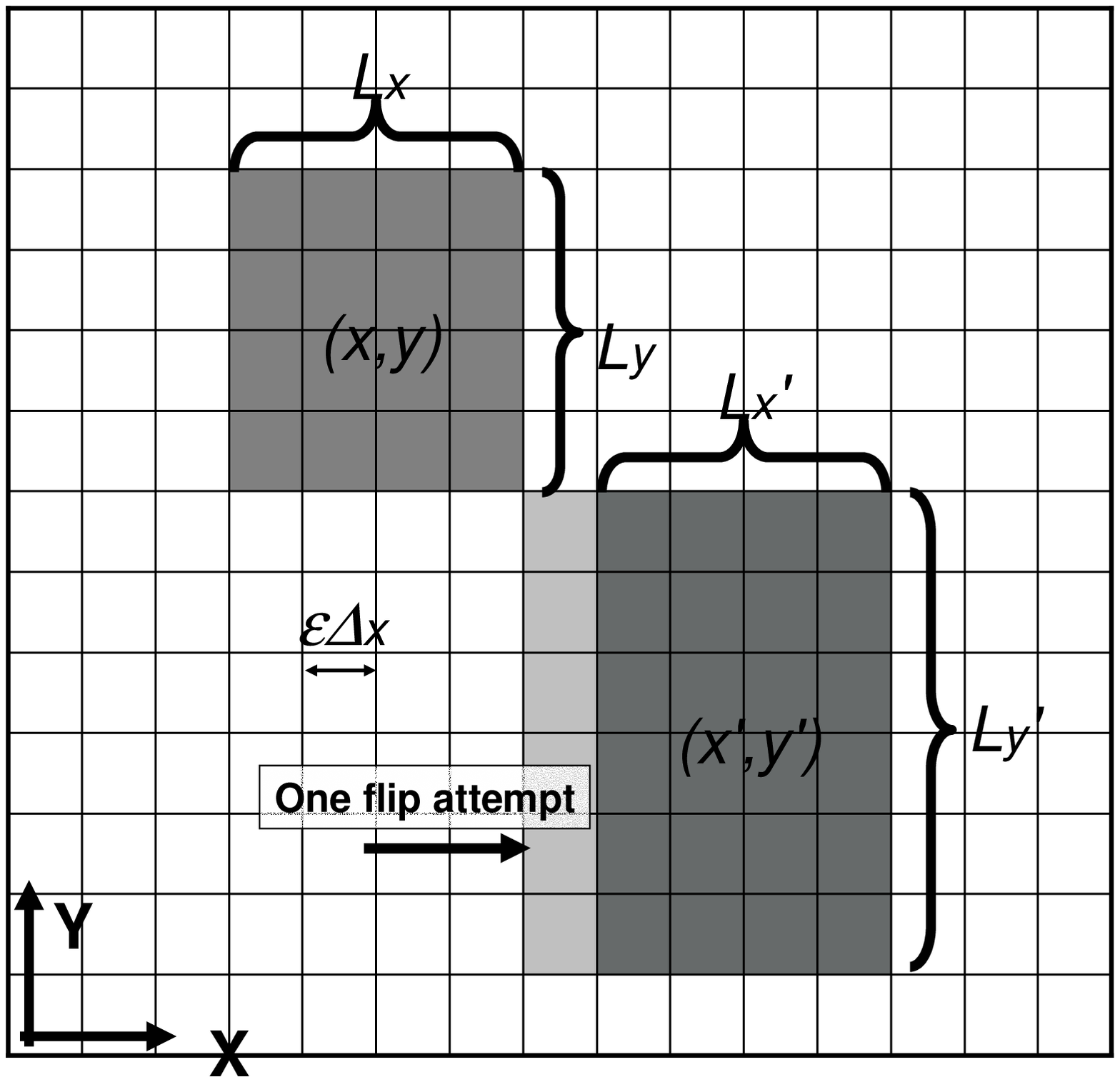}
\caption{2D CPM cell representation. Grey and white colors are used
to indicate cell body and surrounding extracellular matrix
respectively. Cell can grow or shrink in $x$ and $y$ direction by
adding or removing one row (or column) of pixels.
pixels.}\label{figMonte}
\end{center}
\end{figure}
and allow random fluctuation of the dimensions of each
rectangular cell (see Figures \ref{figmultiparticles} and
\ref{figMonte}). Positions and shapes of cells are completely
characterized by a finite set of dynamical variables in the
configuration space: $X=({\bf R}_1, {\bf L}_1, {\bf R}_2, {\bf L}_2,
\ldots, {\bf R}_N, {\bf L}_N)$, where $N$ is the total number of
cells,  ${\bf R}_j$ is a position of center of mass of $j$th cell,
${\bf L}_j$ is the size of  $j$th cell in $D$ spatial dimensions. We
consider $D=1$ and $D=2$ in which case cells are moving over
substrate but results can be extended to the $D=3$ case. Microscopic
description is provided by the multi-cellular probability density
function (PDF) $P(X,t)$ defined as ensemble average $\langle \cdot
\rangle $ over stochastic trajectories $({\bf R}_j'(t),{\bf
L}_j'(t)), \ j=1,\ldots, N$ in the configuration space  $X$:
$P(X,t)=\langle \prod^N_{j=1}\delta({\bf R}_j-{\bf
R}_j'(t))\delta({\bf L}_j-{\bf L}_j'(t))\rangle$  determined by
solution of coupled stochastic equation
\begin{eqnarray}\label{AB1}
\frac{d X}{dt}={\bf  A}(X,c,t)+\hat B(X,c,t)\xi(t),
\end{eqnarray}
and chemotaxis equation for the chemical field $c({\bf r},t)$. Here
${\bf r}$ is the spatial coordinate,  ${\bf A}(X,t)$ is
$2DN$-component vector,  $\hat B(X,t)$ is  $2DN \times 2DN$ matrix
and $\xi(t)$ is $2DN$-component stationary Gaussian stochastic
process with zero correlation time and zero mean
\begin{eqnarray}\label{xidef1}
\langle \xi_j\rangle=0, \ \langle \xi_i(t)\xi_j(t')\rangle=
\delta_{i,j}\delta(t-t'), \ i,j=1,\ldots, 2DN,
\end{eqnarray}
where $\delta_{i,j}$  is a Kronecker's symbol.

Application of the Stratonovich stochastic calculus to the Eq.
(\ref{AB1}) results in a "multi-cellular" Fokker-Planck equation
\cite{Gardiner2004} in configuration space $X$
\begin{eqnarray}\label{FokkerPlanck1}
\partial_t P(X,t)=-\nabla\cdot  \big [{\bf v} P\big ]+
\nabla \cdot \big [ \hat D \cdot \nabla P\big ], \nonumber \\
{\bf v}={\bf  A}-\frac{1}{2}\hat B\cdot \nabla \cdot \hat B^T, \nonumber \\
\hat D=\frac{1}{2}\hat B\cdot \hat B^T,
\end{eqnarray}
where  $\hat D(X,t)$ is a $2DN \times 2DN$ diffusion matrix, $\nabla
=(\partial_{{\bf R}_1},\partial_{{\bf L}_1},\ldots,
\partial_{{\bf R}_N},\partial_{{\bf L}_N})$ is the gradient operator
in $2DN$ dimensions and $\hat B^T$ represents transposed matrix
$\hat B$.

Dynamics of the chemical field $c({\bf r},t)$ is described by a
diffusion equation
\begin{eqnarray}\label{ceqmicroscopic1}
\partial_t c({\bf r},t)=\nabla_{\bf r}  D_c \cdot \nabla_{\bf r} c+
\int A(X,{\bf r},t)P(X,t)dX-\gamma c,
\end{eqnarray}
where $\nabla_{\bf r}$ is the gradient operator for the spatial
coordinate and diffusion coefficient for the chemical field $D_c$ in
general case can depend on ${\bf r}$ and $t$. Term $\int A(X,{\bf
r},t)P(X,t)dX$ describes production of chemical by cells at the rate
of $A(X,{\bf r},t)$ and $\gamma$ is a decay rate of the chemical.
Cells produce chemical into intercellular space through their
membranes. Therefore, in 3D case $A(X,{\bf r},t)$ is nonzero only at
the cellular membrane. However, in 1D and 2D cells are moving on
substrate, while chemical diffuse over entire three dimensional
space so in 1D and 2D cases $A(X,{\bf r},t)$ is nonzero inside
cells.

We assume that ${\bf v}$ has a form of a potential ${\bf
v}=-\beta\nabla \Phi(X,t)$, where $\beta=1/T$ is the inverse
effective temperature $T$ of the chemical fluctuations.
 Multi-cellular Fokker-Planck
equation (\ref{FokkerPlanck1}) is then reduced to multi-cellular
Smoluchowski equation:
\begin{eqnarray}\label{Smoluchowki1}
\partial_t P(X,t)=\nabla \cdot  \hat D \cdot \Big [\nabla P +\beta (\nabla \Phi )P\Big ],
\end{eqnarray}
(More general case of  ${\bf v}$ being any function can be
studied using similar approach.) If we neglect fluctuations of the
cellular size ${\bf L}_j$ and chemotaxis, $c=0$, then Eq.
(\ref{Smoluchowki1}) is similar to the Smoluchowski Eq. for the
Brownian dynamics of colloidal particles (see e.g.
\cite{JonesPusey1991} for a review) and  Eq. (\ref{AB1}) has a form
of the Langevin equation for interacting Brownian particles with
term $\hat B\xi$ representing  thermal forces from solution in
colloids. However, in general case considered here both chemotaxis
and fluctuations of cellular shape are taken into account.
Mechanisms of random fluctuation of cellular shape and
cellular motion are still not completely clear and subject of active
research \cite{DD,CM,AM}.

We impose excluded volume constraint by choosing $\Phi(X,c)=\infty$
if any two cells overlap. We assume in what follows that all direct
interactions between cells is of this type. We also allow indirect
interactions between cells mediated by chemotaxis for which we
choose
\begin{eqnarray}\label{Phidefc1} \left \{\begin{array}{cc}
\Phi(X,c)&=-  \sum\limits_{j=1}^N \chi({\bf L}_j) c({\bf r},t)\Big |_{{\bf r}=X_j}+\Phi(X,c)\Big|_{c=0}
\nonumber\\
& \mbox{if cells do not overlap},  \\
\Phi(X,c)&=\infty
\nonumber\\
& \mbox{if any pair of cells  overlaps},
\end{array} \right.
\end{eqnarray}
where  $\chi({\bf L}_j)$ represents strength of chemotactic
interaction as a function of cellular sizes ${\bf L}_j$ and
$\Phi(X,c)\Big|_{c=0}$ represents chemotaxis-independent terms of
the potential. We assume for simplicity that chemotactic interaction
depends only on gradient of $c({\bf r},t)$ at the center of mass of
each cell. $\Phi(X,c)\Big|_{c=0}$ is also responsible for preserving
cellular shape close to some equilibrium shape. Without this term
shape (size) of each cell would experience unbounded random
fluctuation which is non biological.  We consider specific form of
$\Phi(X,c)$ in Section \ref{section:Macroscopiclimitmasterequation}.

Our main goal is to derive a macroscopic equation describing
dynamics of (total) cell probability density function
\begin{eqnarray}\label{pdef01}
p({\bf r},t)=\sum\limits^{N}_{j=1}p_j({\bf r},t),
\end{eqnarray}
coupled with $c({\bf r},t)$, from microscopic equations
(\ref{ceqmicroscopic1}) and (\ref{Smoluchowki1}). Here $p_j({\bf
r},t)=\int P(X,t)\prod\limits^N_{l=1, \, l\neq j} d{\bf
R}_l\prod\limits^N_{m=1}d{\bf L}_m\Big |_{{\bf R}_j={\bf r}}$ is a
single-cell probability density function of the position of center
of mass. After approximating   $A(X,{\bf
r},t)=a\sum^N_{j=1}\delta({\bf r}-{\bf R}_j)$, $a=const$ and
assuming that $D_c=const$,  Eq. (\ref{ceqmicroscopic1}) is reduced
with the help of Eq. (\ref{pdef01}) to Eq. (\ref{ceqmacroscopic1}).
This approximation is justified because typical diffusion of a
chemical is much faster than cell diffusion  $D_2/D_c\ll 1.$ E.g.
$D_2/D_c\sim 1/40-1/400$ for the cellular slime mold {\it
Dictyostelium} \cite{HoferSherrattMaini1995},   $D_2/D_c\sim 1/30$
and for microglia cells and neutrophils \cite
{LucaChavez-RossEdelstein-KeshetMogilner2003,GrimaPRL2005}.

\section{Microscopic cellular dynamics in Cellular Potts Model}
Stochastic discrete models are used in a variety of problems dealing
with biological complexity. One motivation for this approach is the
enormous range of length scales of typical biological phenomena.
Treating cells as simplified interacting agents, one can simulate
the interactions of tens of thousands to millions of cells and still
have within reach the smaller-scale structures of tissues and organs
that would be ignored in continuum (e.g., partial differential
equation) approaches. At the same time, discrete stochastic models
including the Cellular Potts Model (CPM) can be made sophisticated
enough to reproduce almost all commonly observed types of cell
behavior \cite{Fram1, Cickovski05, Cickovski07, Knewitz06,
Sozinova2006, JRSI}. Recent book \cite{Anderson07} reviews many of
the cell-based models.

The cell-based stochastic discrete CPM, which is an extension of the
Potts model from statistical physics, has become a common technique
for simulating complex biological problems including embryonic
vertebrate limb development \cite{Cickovski05, Newman07}, tumor
growth \cite{Jiang05} and vasculogenesis \cite{Merks}. The CPM can
be made sophisticated enough to reproduce almost all commonly
observed types of cell behavior. It consists of a list of biological
cells with each cell represented by several pixels, a list of
generalized cells, a set of chemical diffusants and a description of
their biological and physical behaviors and interactions embodied in
the effective energy $E$, with additional terms to describe
absorption and secretion of diffusants and extracellular materials.
Distribution of multidimensional indices associated with lattice
cites determines current cell system configuration. The effective
energy of the system, $E$, mixes true energies, like cell-cell
adhesion, and terms that mimic energies, e.g., volume constraint and
the response of a cell to a gradient of an external field (including
chemotactic filed) and area constraint.

\subsection{Cellular Potts Model for cells of rectangular shape}

For simplicity, we use in this paper CPM  with rectangular cellular
shapes. We also assume that all cells have the same type. The results can be extended to the general case of the CPM
with arbitrary cellular shapes. Also, the approach is not limited to
using CPM. For example, one could use microscopic off-lattice models
\cite{NewmanMathBioEng2005,PLoS}, where each cell is represented by
a collection of subcellular elements with postulated interactions
between them.

Notice that reduced representation of cells as fluctuating
rectangles corresponds to intermediate level of description where
fluctuations of cellular shapes are replaced by fluctuations of
cellular sizes. Stochastic Eq. (\ref{AB1}) or Smoluchowski Eq.
(\ref{Smoluchowki1}) coupled with (\ref{ceqmacroscopic1}) can be
used for modeling cell agregation.

In the CPM change of a cell shape evolves according to the classical
Metropolis algorithm based on Boltzmann statistics and the following
effective energy
\begin{equation}
E = E _{ECM}+E _{Adhesion} + E _{Perimeter} + E
_{Field}.\label{eq:2}
\end{equation}

If an attempt to change index of a pixel in a cell leads to energy
change $\Delta E$, it is accepted with the probability
\begin{equation}\label{Phi}
\Phi(\Delta E) = \left \{\begin{array}{cc}1, & \Delta E \leq 0\\
e^{-\beta \Delta E}, &  \Delta E>0, \end{array} \right.
\end{equation}
where $1/\beta=T$ represents an effective boundary fluctuation
amplitude of model cells in units of energy. Since the cells'
environment is highly viscous, cells move to minimize their total
energy consistent with imposed constraints and boundary conditions.
If a change of a randomly chosen pixels' index causes cell-cell
overlap it is abandoned. Otherwise, the acceptance probability is
calculated using the corresponding energy change. The accepted pixel
change attempt results in changing location of the center of mass
and dimensions of the cell.

We consider 2D case with  rectangular shape of each cell with sizes
$ {\bf L}_j=(L_{x,j},L_{y,j})$ and position of center of cellular
mass at ${\bf R}_j=(x_j,y_j)$. Cell motion and changing shape are
implemented by adding or removing a row or column of pixels (see
Figure \ref{figMonte}). We assume that cells can come into direct
contact and that they interact over long distances through
chemotaxis. Term $E _{ECM}$  in the Hamiltonian (\ref{eq:2})
phenomenologically describes net adhesion or repulsion between the
cell surface and surrounding extracellular matrix: $
E_{ECM}=\sum^N_{j=1}2J_{ECM}(L_{x,j}+L_{y,j})$, where $J_{ECM}$ is
the binding energy per unit length of an interface. Term  $E
_{Adhesion}=J_a L_{contact}$ in the Hamiltonian (\ref{eq:2})
corresponds to the   cell-cell adhesion, where $J_a$ is the binding
energy per unit length of an interface and $L_{contact}$ is the
total contact area between cells. Term $E_\mathit{Perimeter}$
defines an energy penalty function for dimensions of a cell
deviating from the target values $L_{T_{x(y)}}$: $ E _{Perimeter} =
\sum^N_{j=1}\lambda_x(L_{x,j}-L_{T_x})^2+\lambda_y(L_{y,j}-L_{T_y})^2$
where $\lambda_x$ and $\lambda_y$ are Lagrange multipliers.  Cells
can move up or down gradients of both diffusible chemical signals (%\emph{i.e.},
\emph{chemotaxis}) and insoluble ECM molecules (%\emph{i.e.},
\emph{haptotaxis}) described by $ E _{Field} = \sum^N_{j=1} \mu \,
c({\bf R}_j,t) L_{x,j}L_{y,j}, $ where $\mu$ is an effective
chemical potential.

In this paper we neglect cell-cell adhesion $J_a=0$ and the
Hamiltonian (\ref{eq:2}) is reduced in 2D to the following
expression
\begin{eqnarray}\label{eq:2reduced}
E(X,t)= \sum\limits^N_{j=1}E({\bf r},{\bf L},t)\Big|_{{\bf r}={\bf R}_j, \ {\bf L}={\bf L}_j}, \nonumber \\
E({\bf r},{\bf L},t)= 2J_{cm}(L_x+L_y)+\lambda_x(L_x-L_{T_x})^2
 \nonumber
\\
\quad +\lambda_y(L_y-L_{T_y})^2+\mu \, c({\bf r}) L_xL_y.
\end{eqnarray}

\subsection{Master equation for discrete cellular dynamics}
We now represent CPM dynamics by using  $P({\bf r},{\bf L},t)$,
probability density for any cell with its center of mass at ${\bf
r}$  to have dimensions ${\bf L}$ at time $t.$ Which means that we
consider one-cell PDF  $P({\bf r},{\bf L},t)$ rather than $N$-cell
PDF $P(X,t)$.   Let $\epsilon \triangle r \times \epsilon \triangle
r$ be the size of a lattice site with $\epsilon \ll 1$ and let
vectors ${\bf e}_{1,2}$ indicate changes in $x$ and $y$ dimensions:
$ {\bf e}_1={\triangle r}(1,0), \ {\bf e}_2=\triangle r (0,1)$. We
normalize the total probability to the number of cells: $\int P({\bf
r},{\bf L},t)d{\bf r}d{\bf L}=N.$ $P({\bf r},{\bf L},t)$ used below
should not be confused with multi-cellular probability density
$P(X,t)$. Those depend on different arguments. Excluded volume
constraint implies that position ${\bf r}'$ and size ${\bf L}'$ of
any neighboring cell should satisfy $ 2|x-x'|\ge L_x+L_x'$, $
2|y-y'|\ge L_y+L_y'$.

Discrete stochastic  cellular dynamics under these conditions is
described by the following master equation
\cite{AlberChenLushnikovNewmanPRL2007}:
%\begin{widetext}
\begin{eqnarray} \begin{array}{cl }
P({\bf r},{\bf L},t+\triangle t) =\sum_{j=1}^2 \{\big
[\frac{1}{2}-\Omega_{j,l}({\bf r}-\frac{\epsilon}{2}{\bf
e}_j,{\bf L}+\epsilon{\bf e}_j; {\bf r},{\bf L}, t) \nonumber  %\\
\end{array}
\end{eqnarray}
\begin{eqnarray} \label{pmasterxL1}\begin{array}{cr cr}
-\Omega_{j,r}({\bf r}+\frac{\epsilon}{2}{\bf e}_j,{\bf L}+\epsilon
{\bf e}_j;{\bf r},{\bf L}, t)-\Omega_{j,l}({\bf
r}+\frac{\epsilon}{2}{\bf e}_j,{\bf L}
 &
 \\
-\epsilon {\bf e}_j; {\bf r},{\bf L}, t) -\Omega_{j,r}({\bf
r}-\frac{\epsilon}{2}{\bf e}_j,{\bf L}-\epsilon {\bf e}_j;{\bf
r},{\bf L}, t)\big ]P({\bf r},{\bf L},t)
 &
 \\
+\Omega_{j,l}({\bf r},{\bf L};{\bf r}+\frac{\epsilon}{2}{\bf e}_j,
{\bf L}-\epsilon {\bf e}_j,t)P({\bf r}+\frac{\epsilon}{2}{\bf
e}_j,{\bf L}-\epsilon {\bf e}_j,t)
 &
 \\
+\Omega_{j,r}({\bf r},{\bf L};{\bf r}-\frac{\epsilon}{2}{\bf e}_j,
{\bf L}-\epsilon {\bf e}_j,t)P({\bf r}-\frac{\epsilon}{2}{\bf
e}_j,{\bf L}-\epsilon {\bf e}_j,t)
 &
 \\
+\Omega_{j,l}({\bf r},{\bf L};{\bf r}-\frac{\epsilon}{2}{\bf
e}_j,{\bf L}+\epsilon {\bf e}_j,t)P({\bf r}-\frac{\epsilon}{2}{\bf
e}_j, {\bf L}+\epsilon {\bf e}_j,t)
&
\\
+\Omega_{j,r}({\bf r},{\bf L};{\bf r}+\frac{\epsilon}{2}{\bf
e}_j,{\bf L}+\epsilon {\bf e}_j,t)P({\bf r}+\frac{\epsilon}{2} {\bf
e}_j, {\bf L}+\epsilon {\bf e}_j,t)  \}. &
\end{array}
\end{eqnarray}
%\end{widetext}
We incorporate dynamics into MC algorithm  by defining the time step
$\triangle t$. Individual biological cells  experience diffusion
through random fluctuations of their shapes. Diffusive coefficient
can be measured experimentally (see e.g. \cite{glazier2000}). We
choose $\triangle t$ to match  experimental diffusion coefficient.
Eq.(\ref{pmasterxL1}) would determine a version of kinetic/dynamic
MC algorithm (see e.g. \cite{kineticMC}) if $\triangle t$ were to be
allowed to fluctuate. For simplicity we assume that $\triangle
t=const$. Also $\Omega_{j,l}({\bf r},{\bf L};{\bf r}',{\bf L}',t)$
and $\Omega_{j,r}({\bf r},{\bf L};{\bf r}',{\bf L}',t)$ denote
probabilities of transitions from a cell of length $L'$ and center
of mass at ${\bf r}'$ to a cell of dimensions $L$ and center of mass
at $\bf r$. Subscripts $l$ and $r$ correspond to transitions by
addition/removal of a  row/colomn of pixels from the rear/lower and
front/upper ends of a cell respectively.

We define  $\Omega_{j,l(r)}({\bf r},{\bf L};{\bf r}',{\bf L}')\equiv
T_{l(r)}({\bf r},{\bf L};{\bf r}',{\bf L}')[1-\phi_{j,l(r)}({\bf
r},\, {\bf L},t)]$ where $T_l(r)({\bf r},{\bf L};{\bf r}',{\bf
L}',t)$  denote probabilities of transitions from a cell of length
${\bf L}'$ and center of mass at ${\bf r}'$ to a cell of dimensions
$\bf L$ and center of mass at $\bf r$ without taking into account
excluded volume principle and cell-cell adhesion. According to the
CPM we have that $T_l({\bf r},{\bf L};{\bf r}',{\bf L}')=T_r({\bf
r},{\bf L};{\bf r}',{\bf L}')= \frac{1}{8}\Phi\Big (E({\bf r},{\bf
L})- E({\bf r}',{\bf L}' ) \Big )$ where the factor of $1/8$ is due
to the fact that there are potentially 8 possibilities for
increasing or decreasing of $L_x$ and $L_y$ (it means that we can
add/romove pixel from any 4 sides of rectangular cell). Second term
$\phi_{j,l(r)}({\bf r},\, {\bf L},t)$ takes into account contact
interactions between cells. It  includes contributions from 3
possible types of stochastic jump processes due to contact
interactions between cells: (a) a cell adheres to another one, (b)
two adhered cells dissociate from each other due to membrane
fluctuations, (c) membranes of two adhered cells are prevented from
moving inside each other (due to excluded volume constraint)
resulting in a negative sign of a contribution to a jump
probability. If neither of these 3 processes happens at given time
step then $\phi_{j,l(r)}({\bf r},\, {\bf L},t)=0$.

\subsection{Macroscopic limit of master equation and mean-field approximation}
\label{section:Macroscopiclimitmasterequation}

Eq. (\ref{pmasterxL1}) is however not closed because one has to know
multi-cellular probability density to determine $\phi_{j,l(r)}({\bf
r},\, {\bf L},t)$. One could use BBGKY-type hierarchy
\cite{McQuarrie2000} similar to the one used in kinetic theory of
gases, which expresses iteratively $n$-cell PDF through $n+1$-cell
PDF with truncation at some order.  This is however extremely
difficult and ineffective for large $n$. Instead, we develop in
Section IV a nonperturbative approach to derivation of Eqs.
(\ref{pottscontinuous4rectangles1Dphi}) and
(\ref{pottscontinuousexcluded2Dphi}).

In previous work
\cite{alber1,alber2,AlberChenLushnikovNewmanPRL2007} we studied a
macroscopic limit $\epsilon\ll 1$ of master Eq. (\ref{pmasterxL1})
by both neglecting contact interactions between cells
\cite{alber1,alber2} and including contact interactions in
mean-field approximation  \cite{AlberChenLushnikovNewmanPRL2007},
which represents simplest closure for BBGKY-type hierarchy. If
contact interactions are neglected, then $\phi_{j,l(r)}({\bf r},\,
{\bf L},t)=0$ and  (\ref{pmasterxL1}) yields in macroscopic limit
$\epsilon \ll 1$ \cite{alber1,alber2}:
\begin{eqnarray}\label{pottscontinuous2D01}
\partial _t P({\bf r},{\bf L},t)=D_2(\nabla^2_{\bf r}
+4\nabla^2_{\bf L})P +8D_2\beta\lambda_x\partial_{L_x}(\tilde
L_xP)\nonumber
\\+8D_2\beta\lambda_y\partial_{L_y}(\tilde L_y P)+D_2\beta
L_xL_y\mu\partial_{\bf r}\big [P\nabla_{\bf r} c\big ],
\nonumber \\
\tilde L_x=\frac{1}{\lambda_x}\big [J_{cm}+\lambda_x (L_x-L_{T_x})
+\frac{1}{2}L_y\mu c({\bf r}) \big ],
 \nonumber \\
\tilde L_y=\frac{1}{\lambda_y}\big [J_{cm}+\lambda_y (L_y-L_{T_y})
+\frac{1}{2}L_x\mu c({\bf r}) \big ],
 \nonumber \\
 D_2=\frac{(\triangle r)^2}{16
\triangle t}, \nabla^2_{\bf r}=\partial_x^2+\partial_y^2, \quad
\nabla^2_{\bf L}=\partial_{L_x}^2+\partial_{L_y}^2.
\end{eqnarray}
Similar Eq. in 1D case (with only coordinate $x$ and length $L_x$
present) was obtained in Ref.  \cite{alber1}.

Mean field approximation assumes decoupling of multi-cellular PDF,
\begin{eqnarray}\label{Pdecouple}
P(X,t)=N^{-N}\prod\limits^N_{j=1}P({\bf r},{\bf L}_j,t)\Big |_{{\bf
r}={\bf R}_j},
\end{eqnarray}
where factor $N^{-N}$ is due to an assumed normalization  $\int
P({\bf r},{\bf L},t)d{\bf r}d{\bf L}=N.$ Mean field approximation is
exact if contact interactions  are neglected. This holds since we
assumed above that chemotaxis depended on averaged density
(\ref{pdef01}) only. Eq. (\ref{Pdecouple}) results in decoupling of
Eq. (\ref{Smoluchowki1}) into independent Eqs. for $P({\bf R}_j,{\bf
L}_j,t)$ for all $j$. This allows direct comparison of Eq.
(\ref{pottscontinuous2D01}) with Eq. (\ref{Smoluchowki1}) and
yields that diffusion matrix has a diagonal form  with main diagonal
$D_2(1,1,4,4,1,1,4,4 \ldots )$ in 2D. For 1D and 3D cases numbers
$1$ and $4$ repeat themselves with period $1$ and $3$, respectively.

 Further comparison of  (\ref{Smoluchowki1}), (\ref{Phidefc1}) and
 (\ref{pottscontinuous2D01}) leads to expression $\chi({\bf L}_j)=\ L_{x,j}L_{y,j} \mu$ and
\begin{eqnarray}\label{Phidefc2}
\Phi(X,t)=E(X,t),
\end{eqnarray}
where $E(X,t)$ is given by (\ref{eq:2reduced}).

Taking into account contact cell-cell interaction (excluded volume)
yields that Eq. (\ref{Pdecouple}) is not exact any more. Also,
potential $\Phi(X,t)$ is given by (\ref{Phidefc2}) only if cells do
not overlap. Otherwise $\Phi(X,t)=\infty$ according to
(\ref{Phidefc1}).

\subsection{Boltzmann-like distribution and macroscopic equation for cellular density}

From Eqs. (\ref{eq:2reduced}), (\ref{Phi}) and
(\ref{pottscontinuous2D01}) it follows that  typical fluctuations of
cell sizes $\tilde L_{x(y)}$ are determined by
$\beta\lambda_{x(y)}\tilde L_{x(y)}^2\sim 1$. Suppose $x_0$ and
$y_0$ are typical scales of $P({\bf r},{\bf L},t)$ with respect to
$x$ and $y$. We assume that $ \beta x_0^2\lambda_x \gg 1$ and $\quad
\beta y_0^2\lambda_y \gg 1$, meaning that $x_0\gg \tilde L_x, \
y_0\gg \tilde L_y$. We also assume that chemical field $c({\bf
r},t)$ is a slowly varying function of ${\bf r}$ on the scale of
typical cell length meaning that $ x_c/L_x\gg 1, \quad y_c/L_y\gg
1,$ where $x_c$ and $y_c$ are typical scales for variation of
$c({\bf r},t)$ in $x$ and $y$.  We also make an additional
biologically relevant assumption that $ 4\lambda_x\lambda_y\gg
\mu^2c({\bf r},t)^2$ meaning that change of typical cell size  due
to chemotaxis $\delta L^{(chemo)}_{x(y)}$ is small $|\delta
L^{(chemo)}_{x(y)}|\ll L^{(min)}_{x(y)}.$

If all these conditions are satisfied then we found by using both
solutions of Eq. (\ref{pottscontinuous2D01}) and MC simulations of
CPM   with  general initial conditions,   that PDF $P({\bf r},{\bf
L},t)$ quickly converges in $t$  at each spatial point ${\bf r}$ to
the following Boltzmann - like form
\begin{eqnarray}\label{pxdef1}
P({\bf r},{\bf L},t)=P_{Boltz}({\bf r},{\bf L})p({\bf r},t),
\end{eqnarray}
where
\begin{eqnarray}\label{Boltzmann2d}
P_{Boltz}({\bf r},{\bf L})=\frac{1}{Z({\bf r})}\exp(-\beta \triangle
E_{length})
\end{eqnarray}
  is a Boltzmann - like
distribution depending on ${\bf r}$ and $t$ only through $c({\bf
r},t)$,
\begin{eqnarray}\label{dElengthdef2d}
\triangle E_{length}=E({\bf r},{\bf L})-E_{min}
\nonumber\\
=\lambda_x L_x^{\prime 2}+\lambda_y L_y^{\prime 2}+ L'_x L'_y\mu
c({\bf r}),
%
%\nonumber \\
%
{\bf L}'={\bf L}-{\bf L}^{(min)}.
\end{eqnarray}
Here $E_{min}=E({\bf r},{\bf L}^{(min)})$  is the minimal value of
(\ref{eq:2reduced}) achieved at ${\bf L}={\bf L}^{(min)}$:
\begin{eqnarray}\label{Lmin}
E_{min}=E({\bf r},{\bf L}^{(min)}),
\nonumber \\
L_{x}^{(min)}=\frac{-4\lambda_y(J_{cm}-\lambda_xL_{T_x})+2(J_{cm}-\lambda_yL_{T_y})\mu
c({\bf r})}{4\lambda_x\lambda_y-\mu^2c({\bf r})^2},
\nonumber \\
L_{y}^{(min)}=\frac{-4\lambda_x(J_{cm}-\lambda_yL_{T_y})+2(J_{cm}-\lambda_xL_{T_x})\mu
c({\bf r})}{4\lambda_x\lambda_y-\mu^2c({\bf r})^2}
\nonumber\\
~
\end{eqnarray}
and  $ Z({\bf r},t)=(2\epsilon\triangle r)^2\sum_{{\bf
L}}\exp(-\beta \triangle E_{length})\simeq
\frac{2\pi}{\beta\sqrt{4\lambda_x\lambda_y-\mu^2c({\bf r},t)^2}} $
is an asymptotic formula for a partition function as $\epsilon\to
0$. (See \cite{alber1} for details about convergence rate for the
case without contact interactions)

Also, under these conditions we can use Eq. (\ref{pxdef1}) to
integrate Eq. (\ref{pottscontinuous2D01}) over $\bf L$ which results
in the following evolution equation for the the cellular probability
density $p({\bf r},t)=\int (P({\bf r},{\bf L},t)d{\bf L}$
\cite{alber1,alber2}:
\begin{eqnarray}\label{pottscontinuous3}
\partial _tp=D_2\nabla^2_{\bf r} p-\chi_0\nabla_{\bf r} \cdot \big [p\, \nabla_{\bf
r}c({\bf r},t)\big ],
\end{eqnarray}
where
 $ \chi_0=-D_2\mu \beta L_{x}^{(0)}L_{y}^{(0)}$ and
\begin{eqnarray}\label{Lminreduced}
L_{x}^{(0)}=L_{T_{x}}-J_{cm}/\lambda_{x}, \nonumber \\
L_{y}^{(0)}=L_{T_{y}}-J_{cm}/\lambda_{y}
\end{eqnarray}
correspond to $L_{x(y)}^{(min)}$ from (\ref{Lmin})  provided we
neglect chemotaxis.

Eq. (\ref{pottscontinuous3}) together with $(\ref{ceqmacroscopic1})$
form a closed set which coincides with the classical Keller-Segel
system \cite{KS}. It has a finite time singularity (collapse) and
was extensively used for modeling aggregation of bacterial colonies
\cite{HerreroVelazquez1996,BrennerConstantinKadanoff1999}. We show
below that near singularity contact interactions between cells could
prevent collapse.

Eqs. similar to  $(\ref{pottscontinuous3})$ and
$(\ref{ceqmacroscopic1})$  for the case of contact interactions
(excluded volume constraint) has been obtained in mean-field
approximation \cite{AlberChenLushnikovNewmanPRL2007}. These Eqs.
significantly slow down collapse in comparison with
$(\ref{pottscontinuous3})$ and $(\ref{ceqmacroscopic1})$. They still
have collapsing solutions if initial density is not small. (These
Eqs. are applicable only for small densities.)

\section{Beyond mean-field approximation and regularization of collapse}
Main purpose of this paper is to derive macroscopic Eqs. which do
not have collapse (blow up of solutions in finite time) for
arbitrary initial densities and are in good agreement with
microscopic stochastic simulations for large cellular densities.
This requires one to go beyond mean-field approximation.

We conclude from the previous section that random changes of
cellular lengths result in random walks of centers of mass of cells
during the time between cell "collisions". Significant
simplification in comparison with Eq. (\ref{FokkerPlanck1}) is that
we have now explicit dependence on spatial coordinate $\bf r$ but
not on $\bf L$. We use below notation $(L_{x}^{(0)},L_{y}^{(0)})$
for average size of a cell neglecting change of that size due to
chemotaxis. For CPM $(L_{x}^{(0)},L_{y}^{(0)})$ are given by
(\ref{Lminreduced}). If we neglect chemotaxis (i.e. set $c=0$) then
during time between collision  cell probability density $p({\bf
r},t)$ is described by linear diffusion  Eq. which follows from Eq.
(\ref{pottscontinuous3}):
\begin{eqnarray}\label{diffusionlinear1a}
\partial _tp=D_2\nabla^2_{\bf r} p.
\end{eqnarray}
In a similar way, two-cell probability density $p({\bf r}_1,{\bf r}_2,t)$
is described by linear diffusion  for two independent variables ${\bf r}_1,{\bf r}_2$:
\begin{eqnarray}\label{diffusionlinear2}
\partial _tp=D_2(\nabla^2_{{\bf r}_1}+\nabla^2_{{\bf r}_2}) p.
\end{eqnarray}
 $p({\bf r}_1,{\bf r}_2,t)$ represents probability density of a cells
 $1$ and $2$ having centers of mass at ${\bf r}_1$ and  ${\bf r}_2$ at time $t$, respectively.
 After making change of variables
\begin{eqnarray}\label{rR1}
 {\bf r}={\bf r}_1-{\bf r}_2, \quad {\bf R}=({\bf r}_1+{\bf r}_2)/2,
  \end{eqnarray}
where variable ${\bf r}$ describes relative motion of cells and
variable ${\bf R}$ describes motion of "center of mass" of two
cells, Eq. (\ref{diffusionlinear2}) takes the following form
\begin{eqnarray}\label{diffusionlinear2rel}
\partial _tp=2D_2\nabla^2_{{\bf r}}p+\frac{D_2}{2}\nabla^2_{{\bf R}} p.
\end{eqnarray}
Each collision involving cell $1$ or $2$ modifies both $p({\bf
r},t)$ and $p({\bf r}_1,{\bf r}_2,t)$. In other words, it effects
random walk of each colliding cell.

We describe first the effect of collisions due to excluded volume
constraint between cells in 1D case. Consider a pair of neighboring
cells which in 1D always remain neighbors. We assume that at the
time of collision two colliding cells have the same size. Generally
this is not true because sizes of cells continuously fluctuate with
lengthscale $\delta L_x\sim 1/(\beta^{1/2}\lambda_x^{1/2}).$
However, we assume as before that these fluctuation are small
$|\delta L_x| \ll L_x^{(0)}$  which justifies our approximation.
Collision is defined as two cells being in direct contact at given
moment of time and one of them trying to penetrate into another.
Collison is prevented by excluded volume constraint. In the
continuous limit $\epsilon\to 0$ each collision takes
infinitesimally small time. After collision cells move away from
each other so they are not in direct contact any more. Instead of
explicitly describing each of these collisions we use an assumption
that two cells are identical and view each collision as exchange of
positions of two cells \cite{Rost1984,BodnarVelazquez2005}. From such point of view cells
do not collide at all but simply pass through each other. They both
experience random walk as point objects (cells) without collisions
according to Eqs. (\ref{diffusionlinear1a}) and
(\ref{diffusionlinear2rel}) in the domain free from other cells
(free domain). (The volume of such free domain per cell, which has
dimension of length in 1D, is  on average $(1-L_x^{(0)}
p(x,t))/p(x,t)$.) This means that we are considering collective
diffusion \cite{JonesPusey1991} of cellular density instead of
trajectories of individual cells. In contrast, self-diffusion
describes mean-square displacement of an individual cell as a
function of time \cite{JonesPusey1991}. This approach might be
important for describing propagation of one cell type through space
occupied by cells of another type. In this paper we consider only
collective diffusion.

While trajectories of cells in free domain are continuous, positions
of cells in  physical space change instantaneously at each collision
by $L_x$ for the cell colliding from the left and by $-L_x$ for the
cell colliding from the right. The effective rate of cell diffusion
is enhanced as free space becomes smaller with growth of cellular
volume fraction. Let's assume that at the initial time $t=t_0$
centers of mass of two cells are separated by an average distance
$1/p(x,t)$ and that these two cells collide for the first time
(meaning that previous collisions of each cell involved collision
with cells other than these two). This yields that their centers of mass
are separated by distance $L_x^{(0)}$ at $t=t_0.$ If moving
reference frame is set at one of the cells then other will
experience random walk with doubled diffusion coefficient $2D_2$ (as
seen from Eq. (\ref{diffusionlinear2rel})), where $D_2$ is a
diffusion coefficient of each cell in the stationary reference
frame. Relative motion of two cells in moving reference frame
corresponds to random walk of a point-wise cell with diffusion
coefficient $2D_2$. Consider continuous random walk. Number of
returns to the initial position $x=L_x^{(0)}$ during any finite time
interval after $t_0$ is infinite meaning that number of collisions
between given pair of cells in physical space is infinite. However,
successive collisions effectively cancel each other since they
change positions of cells by $\pm L_x^{(0)}.$ What really matters is
whether total number of collision is even or odd. For even number of
collisions total contribution of collisions is zero. While for odd
number of collisions contribution to the flux of probability for the
left cell (at time $t_0$) $\propto L_x^{(0)}$ and for the right cell
$\propto-L_x^{(0)}.$ Because total number of collisions is infinite,
the probabilities to have even or odd number of collisions are equal
to $1/2.$ While the average distance between center of mass of cells
is $1/p(x,t)$, the average distance between boundaries of two
neighboring cells  in physical domain is
\begin{eqnarray}\label{deltax1Dfree}
\triangle x=1/p(x,t)-L_x^{(0)}.
\end{eqnarray}
When separation between surfaces of two cells after
collision reaches  $\triangle x$ from (\ref{deltax1Dfree}) we determine that "extended
collision'' between the pair of cells is over. Namely, two cells are
not closer to each other than to other neighboring cells any more.
Therefore, probability of them colliding with each other is not
higher any more than probability of them colliding with other
cells. This extended collision includes infinitely many
"elementary" collisions but its final contribution depends only on
whether the total number of such collision is even or odd.

To find average time of extended collision we solve diffusion Eq. in
moving frame
\begin{eqnarray}\label{diffusionlinear1}
\partial _tp_m=2D_2\nabla^2_{x} p_m
\end{eqnarray}
with reflecting boundary condition $\partial_x p_m(L_x^{(0)},t)=0$
at $x=L_x^{(0)}$ and absorbing boundary condition
$p_m(L_x^{(0)}+\triangle x,t)=0$ at $x=L_x^{(0)}+\triangle x$. Reflecting boundary condition means that
cell does not cross point $x=L_x^{(0)}$. Instead of
crossing $x=L_x^{(0)}$ cells exchange positions at each collision.
Absorbing boundary condition corresponds to the "escape point" of cell
from extended collision. Initial condition is
$p_m(x,t_0)=\delta(x-L_x^{(0)})$ which is defined by initial zero
distance between surfaces of two cells. Solution of the Eq.
(\ref{diffusionlinear1}) with these initial and boundary conditions
results in calculation of the  mean first-passage time (escape time)
$T_{esc}$, which is equal to the extended collision time in our
case. To find $T_{esc}$ we use a backward Fokker-Planck Eq.
\begin{eqnarray}\label{backwardfokkerplank0}
\partial _{t_1}p(x_2,t_2|x_1,t_1)=U'(x_1)\partial_{x_1}p(x_2,t_2|x_1,t_1)
\nonumber\\
-D_0\nabla^2_{x_1} p(x_2,t_2|x_1,t_1)
 \end{eqnarray}
(see e.g. \cite{Gardiner2004}),  where $p(x_2,t_2|x_1,t_1)$ is the
conditional probability density for transition from $(x_1,t_1)$ to
$(x_2,t_2)$. $D_0$ is the diffusion coefficient and $U(x)$ is an
arbitrary potential. In 1D case $U(x)=0$ and $D_0=2D_2$.

For convenience of the reader we provide here a derivation of
$T_{esc}$ which is similar to the one from Ref. \cite{Gardiner2004}.
Probability that cell remains inside an interval $(0,\triangle x)$
at time $t$, provided it was located at $x$ at $t=t_0$, is given by
$G(x,t)\equiv \int^{\triangle x}_0 p(x',t|x,t_0)dx'$. Using the
stationarity of the random walk $p(x',t|x,t_0)=p(x',0|x,t_0-t)$ we
obtain from (\ref{backwardfokkerplank0}) that
\begin{eqnarray}\label{backwardfokkerplank1}
\partial _tG=-U'(x)\partial_xG+D_0\nabla^2_{x} G.
\end{eqnarray}
Mean escape time $T_{esc}(x)$ for a cell located at $x$ at $t=t_0$
is
\begin{eqnarray}\label{Tescdef1}
T_{esc}(x)=-\int\limits^{\infty}_{t_0}(t-t_0)\partial_tG(x,t)dt=\int\limits^{\infty}_{t_0}G(x,t)dt.
 \end{eqnarray}
Integrating Eq. (\ref{backwardfokkerplank1}) over $t$ from $t_0$ to
$\infty$ and using initial normalization $G(x,t_0)=1$ results in
\begin{eqnarray}\label{Tescode1}
-U'(x)\partial_xT_{esc}(x)+D_0\nabla^2_{x}T_{esc}(x)=-1.
\end{eqnarray}
Reflecting and absorbing boundary conditions for $p$ result in
similar boundary conditions for $T_{esc}(x):$
\begin{eqnarray}\label{Tescbond1}
\partial_xT_{esc}(x)\Big|_{x=L_x^{(0)}}=0, \quad  T_{esc}(x)\Big|_{x=L_x^{(0)}+\triangle x}=0,
\end{eqnarray}
which allows us to solve boundary value problem
(\ref{Tescode1}),(\ref{Tescbond1}) explicitly:
\begin{eqnarray}\label{Tescsol1}
T_{esc}(x)=D_0^{-1}\int\limits^{L_x^{(0)}+\triangle x}_{x}\exp{\big [  D_0^{-1} U(x') \big]} dx'
\nonumber\\
\times \int \limits^{ x'}_{L_x^{(0)}}\exp{\big [  -D_0^{-1} U(x'') \big]} dx''.
\end{eqnarray}
Initial condition implies that $T_{esc,1}=T_{esc}(L_x^{(0)})$ and
yields extended collision time in 1D:
\begin{eqnarray}\label{Tescs1D}
T_{esc,1}=\frac{(\triangle x)^2}{4D_2}.
\end{eqnarray}
Consider mean square displacement of position of each cell during
extended collision in the stationary frame. We use notations
$(\triangle x)^2_1$ and $(\triangle x)^2_2$ for  mean square
displacement of  cells $1$ and $2$, respectively,   in the
stationary frame. Using (\ref{rR1}) we obtain that $(\triangle
x)^2_1+(\triangle x)^2_2=\frac{(\triangle x)^2}{2}+2(\triangle
X)^2$, where $X={\bf R}$ in 1D. Because of the symmetry between
cells $1$ and $2$ the following relation holds $(\triangle
x)^2_1=(\triangle x)^2_2=\frac{(\triangle x)^2}{4}+(\triangle X)^2$.
Center of mass of two cells experiences random walk with diffusion
coefficient $D_2/2$ (see Eq. (\ref{diffusionlinear2rel})) and
$(\triangle X)^2=T_{esc,1}D_2=\frac{(\triangle x)^2}{4},$
$(\triangle x)^2_1=(\triangle x)^2_2=\frac{(\triangle x)^2}{2}$.

Now recall that with probability $1/2$ cells exchange positions
during extended collision meaning that  in expression for
$(\triangle x)^2_{1,2}$ with probability $1/2$ term $(\triangle
x)^2$ should be replaced by $(2L_x^{(0)}+\triangle x)^2$ resulting
in the total mean square displacement for cell $1$ or $2$
\begin{eqnarray}\label{deltax1tot}
 (\triangle x)^2_{ total}=\frac{1}{2}\Big (\frac{(\triangle x)^2}{2}+\frac{(2L_x^{(0)}+\triangle x)^2}{2}\Big ).
\end{eqnarray}
From Eqs. (\ref{deltax1Dfree}),(\ref{Tescs1D}) and
(\ref{deltax1tot}) we obtain effective (nonlinear) diffusion
coefficient for the cell probability density
\begin{eqnarray}\label{diff1DeffectiveNoN}
D_{1, effective}=\frac{ (\triangle x)^2_{ total}}{2T_{esc,1}}=D_2\Big [ \frac{1+(L_x^{(0)})^2 p^2}{(1- L_x^{(0)} p)^2}\Big ].
\end{eqnarray}
In simulations described below we often use not very large number of
cells $N$. In case of finite $N$ each given cell can collide with
$N-1$ remaining cells which gives extra factor $(N-1)/N$ and allows
one to rewrite  the effective diffusion coefficient
(\ref{diff1DeffectiveNoN}) as follows
\begin{eqnarray}\label{diff1Deffective}
D_{1, effective}=D_2\Big [ \frac{1+(1-N^{-1})(L_x^{(0)})^2 p^2}{(1- (1-N^{-1})L_x^{(0)} p)^2}\Big ].
\end{eqnarray}
The effective diffusion (\ref{diff1Deffective}) results in 1D
nonlinear diffusion  Eq.:
\begin{eqnarray}\label{pottscontinuous4rectangles1D}
\partial _tp=D_2\nabla_{\bf r} \cdot \Big [ \frac{1+(1-N^{-1})(L_x^{(0)})^2 p^2}{(1- (1-N^{-1})L_x^{(0)} p)^2}\nabla_{\bf r} p\Big ]
\nonumber\\
-\chi_0\nabla_{\bf r} \cdot \big [p\, \nabla_{\bf
r}c({\bf r},t)\big ].
\end{eqnarray}
Here we added chemotaxis term from Eq. (\ref{pottscontinuous3})
which is well justified provided $x_c\gg \triangle x$
(see also previous section).

Now consider 2D case. First consider cells with disk-shaped form instead of cells of
rectangular shapes (i.e. cellular shapes fluctuating around a disk)
with average diameter $L_x^{(0)}$. Assume, similar to 1D case, that
$t_0$ is the time of the first collision between two given cells.
Previous collisions of each of these cells involved cells other then
these two. We also average over angles relative to the center of one
of these two cells which implies  that cells collide at time $t=t_0$
with equal probability at every angle. In that approximation there
is rotational symmetry in the moving frame and all variables depend
only on the radial variable $r\equiv|{\bf r}|$ but do not depend on
angular variables. Change of variables in Fokker-Planck Eq.
$\partial _tp=2D_2\nabla^2_{\bf r} p$ results in
\begin{eqnarray}\label{fokkerplank2D3D}
\partial _t{\tilde p}=-2D_2\partial_r\frac{\tilde p}{r}+2D_2\nabla^2_{r} \tilde p,
\end{eqnarray}
where  $\tilde p\equiv r p.$ Eq. (\ref{fokkerplank2D3D}) is
equivalent to the 1D Fokker-Planck Eq. with potential
$U(r)=-2D_2\log(r)$. Backward Fokker-Planck Eq.
(\ref{backwardfokkerplank0}) with $x=r$, and the same $U(r)$ yields
in a way similar to Eqs.
(\ref{backwardfokkerplank1})-(\ref{Tescs1D}), an extended collision
time (mean escape time)
\begin{eqnarray}\label{Tescs2D}
T_{esc,2}=\frac{\triangle x(2L_x^{(0)}+\triangle x)}{8D_2}-
\frac{\big [L_x^{(0)}\big ]^2}{4D_2}\log{\left (1+\frac{\triangle x}{L_x^{(0)}}\right )}.
\end{eqnarray}
We assume again that during extended collision time cells on average
span entire space, i.e.
\begin{eqnarray}\label{detaxdefD2D3}
\frac{\pi}{4}(L_x^{(0)}+\triangle x)^2=p^{-1}.
\end{eqnarray}
Using Eqs. (\ref{deltax1tot}), (\ref{Tescs2D}) and
(\ref{detaxdefD2D3}) we obtain effective diffusion coefficient in 2D
(here and below the average disk diameter is $L\equiv L_x^{(0)}$):
\begin{eqnarray}\label{diff2DeffectiveNoN}
D_{2, effective}=\frac{ (\triangle x)^2_{ total}}{4T_{esc,2}}
\nonumber\\
=D_2 \Big [ \frac{1+\pi L^2 p}{1- \pi L^2 p+\pi L^2 p\log(\pi L^2 p)}\Big ],
\end{eqnarray}
which after modification to include effect of a finite $N$ similar
to the one used in 1D case, results in
 \begin{widetext}
\begin{eqnarray}\label{diff2Deffective}
D_{2, effective}=D_2 \Big [ \frac{1+(1-N^{-1})\pi L^2 p}{1- (1-N^{-1})\pi L^2 p+(1-N^{-1})\pi L^2 p\log(\pi L^2 p)}\Big ],
\end{eqnarray}
Eq. (\ref{diff2Deffective}) yields the following nonlinear diffusion
Eq. for cells fluctuating around disk-shaped form
\begin{eqnarray}\label{pottscontinuous4spheres}
\partial _tp=D_2\nabla_{\bf r} \cdot \Big [ \frac{1+(1-N^{-1})\pi L^2 p}{1- (1-N^{-1})\pi L^2 p+(1-N^{-1})\pi L^2 p\log(\pi L^2 p)}\nabla_{\bf r} p\Big ]-\chi_0\nabla_{\bf r} \cdot \big [p\, \nabla_{\bf
r}c({\bf r},t)\big ].
\end{eqnarray}
Notice that both 1D effective diffusion coefficient
(\ref{diff1Deffective}) and 2D effective diffusion coefficient
(\ref{diff2Deffective}) depend only on the volume fraction $\phi$
($\phi=L_x^{(0)}p$ in 1D and $\phi=\pi L^2 p$ in 2D). Based on that
we propose that the  effective diffusion coefficient  for 2D
rectangles also depends only on the volume fraction
$\phi=L_x^{(0)}L_y^{(0)}p$, which results in the  nonlinear
diffusion Eq. for cells of rectangular shape
\begin{eqnarray}\label{pottscontinuous4rectangles}
\partial _tp=D_2\nabla_{\bf r} \cdot \Big [ \frac{1+(1-N^{-1}) L_x^{(0)}L_y^{(0)}  p}{1- (1-N^{-1})L_x^{(0)}L_y^{(0)} p+(1-N^{-1})L_x^{(0)}L_y^{(0)} p\log(L_x^{(0)}L_y^{(0)} p)}\nabla_{\bf r} p\Big ]-\chi_0\nabla_{\bf r} \cdot \big [p\, \nabla_{\bf
r}c({\bf r},t)\big ].
\end{eqnarray}
\end{widetext}
Numerical simulations in the next section confirm that Eq.
(\ref{pottscontinuous4rectangles}) agrees very well with the MC
simulations of microscopic dynamics.

In macroscopic limit $N \gg 1$ factor $ (1-N^{-1})$ is replaced by
$1$ in Eqs. (\ref{pottscontinuous4rectangles1D}),
(\ref{pottscontinuous4spheres}) and
(\ref{pottscontinuous4rectangles}), which yields Eqs.
(\ref{pottscontinuous4rectangles1Dphi}) and
(\ref{pottscontinuousexcluded2Dphi}).

If volume fraction $\phi\to 1$ then nonlinear diffusion in Eqs.
(\ref{pottscontinuous4rectangles1Dphi}) and
(\ref{pottscontinuousexcluded2Dphi}) diverges. Thus we do not expect
any blow up of the solution contrary to Eq.
(\ref{pottscontinuous3}). Global existence of Eqs.
(\ref{pottscontinuous4rectangles1Dphi}) and
(\ref{pottscontinuousexcluded2Dphi}) together with Eq.
(\ref{ceqmacroscopic1}) can be studied in a way similar to
\cite{HillenWang2007}.

\section{Numerical results and application to vasculogenesis}
\subsubsection{One dimensional cell motion}\label{nodyn}
\begin{figure}
\begin{center}
\includegraphics [scale=0.75]{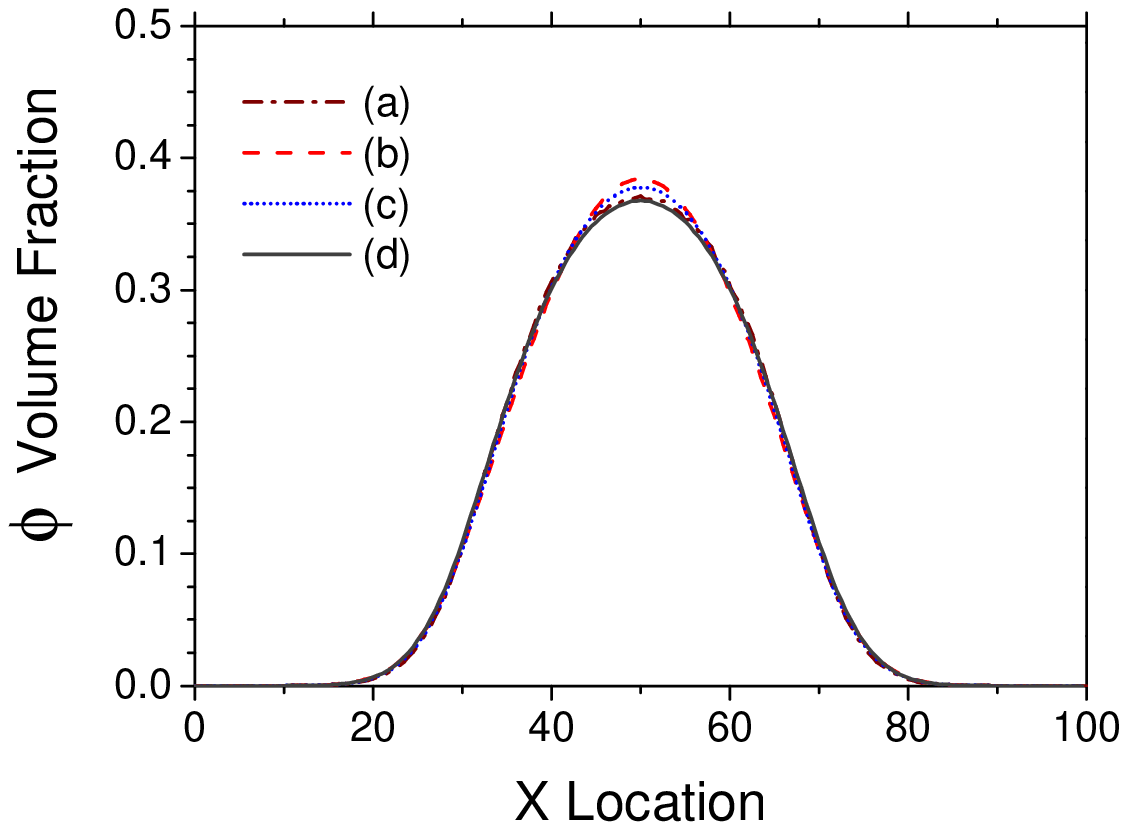}
\end{center}
\caption{Volume fraction $\phi=L_x^{(0)} p(x,t_{end})$ for the 1D
cellular motion as a function of $x$. Curve (a): Monte Carlo
simulations of the CPM. Curve (b): solution of the Eq.
(\ref{pottscontinuous3}). Curve (c):  solution of the Eq.
(\ref{pottscontinuous1DPercus}). Curve (d) solution of the Eq.
(\ref{pottscontinuous4rectangles1D}).}\label{fig:fig1d}
\end{figure}

Figure \ref{fig:fig1d} demonstrates simulation of 1D motion of cells
represented in the form of fluctuating rods. Numerical solution of
the continuous model was obtained using a pseudo-spectral scheme.
For both the CPM simulation and numerics of the continuous model we
used periodic boundary conditions, simulation time was $t_{end}=200$
and values of other parameters were chosen as follows: $\triangle
r=1$, $L_{T_x}=L_{T_x}=3$, $\lambda_x=\lambda_y=1.5$,
$\epsilon=0.01$ $J_{cm}=2$, $\beta=15$, $N=8$, $\mu=0$. Simulation
was performed on the spatial domain $0 \le x < 100$ and the initial
cell density distribution was $p({\bf r},0)= k_0e^{-((x-50)/10)^4}$
with $k_0$ determined by normalization $N=8$. Figure \ref{fig:fig1d}
shows simulations of the CPM (curve a), numerical solutions of the
macroscopic model without excluded volume interactions
(\ref{pottscontinuous3}) (curve b), macroscopic Eq.
(\ref{pottscontinuous1DPercus}) (curve c), and macroscopic Eq.
(\ref{pottscontinuous4rectangles1D}) (curve d).
Here Eq.
\begin{eqnarray}\label{pottscontinuous1DPercus}
\partial _tp=D_2\nabla_{\bf r} \cdot \Big [ \frac{1}{(1- (1-N^{-1})L_x^{(0)} p)^2}\nabla_{\bf r} p\Big ]
\nonumber\\
-\chi_0\nabla_{\bf r} \cdot \big [p\, \nabla_{\bf
r}c({\bf r},t)\big ],
\end{eqnarray}
was derived from the equation of state for the  1D hard rod fluid
\cite{PercusJStatPhys1976} which allows one to determine collective
diffusion coefficient from static structure factor and
compressibility (see e.g. \cite{JonesPusey1991,Gomer1990}).

Figure \ref{fig:fig1d} demonstrates that solution of the Eq.
(\ref{pottscontinuous4rectangles1D}) is in much better agreement
with CPM than both Eqs. (\ref{pottscontinuous3}) and
(\ref{pottscontinuous1DPercus}). While  the difference between the
CPM simulation and solutions of the Eq. (\ref{pottscontinuous3}) and
Eq. (\ref{pottscontinuous1DPercus}) is small but clearly exceeds the
error in MC simulations. Difference between MC simulation and
solution of the Eq. (\ref{pottscontinuous4rectangles1D}) is within
an accuracy of the MC simulations.

Difference between CPM and Eq.
(\ref{pottscontinuous1DPercus}) is due to the fact that the equation of state
 for the 1D hard rod fluid was calculated in Ref.
\cite{PercusJStatPhys1976} from grand canonical partition function
\cite{LandauLifshitzV5} while diffusion of cells is a nonequilibrium
phenomenon resulting in the corrections to the equilibrium partition
function. Numerous attempts have been made to describe dynamics of
interacting Brownian particles (see \cite{KawasakiJStatMech2008} and
reference there in). Note also that difference between CPM  and Eq.
(\ref{pottscontinuous3}) results is not so dramatic in 1D as in 2D
because in 1D  Keller-Segel model does not support collapse
\cite{BrennerConstantinKadanoff1999}.

\subsubsection{Two dimensional cell motion with chemotaxis}

\begin{figure}
\begin{center}
\includegraphics [scale=0.6]{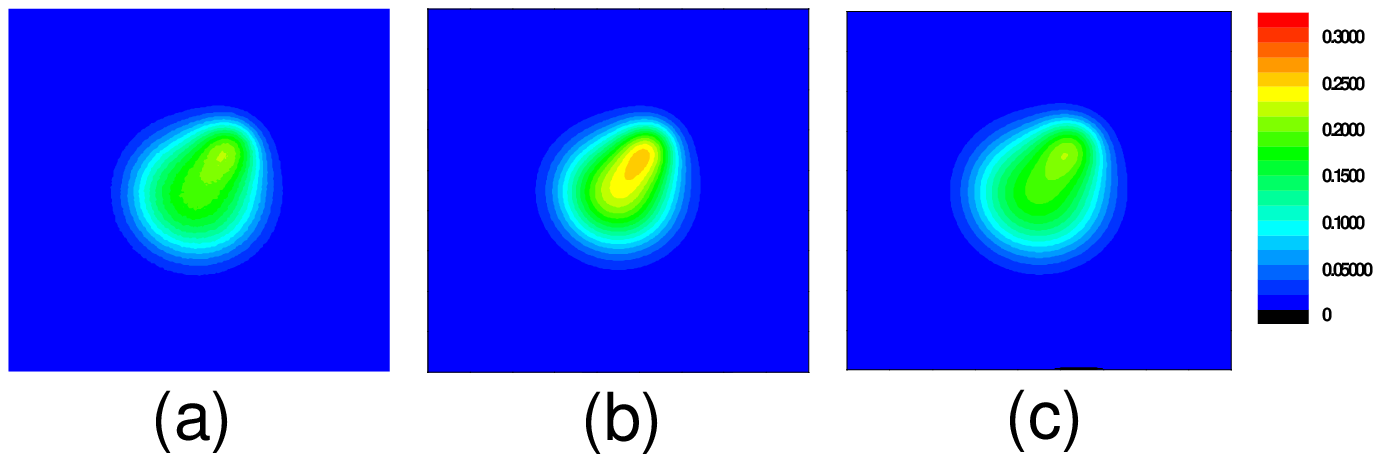}
\end{center}
\caption{$p(x,y,t)$ in 2D as a function of $(x,y)$ for (a)
Monte Carlo simulation of CPM, (b)  Eq. (\ref{pottscontinuous3})
and (c)  Eq. (\ref{pottscontinuous4rectangles}). }\label{figTwoD}
\end{figure}

\begin{figure}
\begin{center}
\includegraphics [scale=0.75]{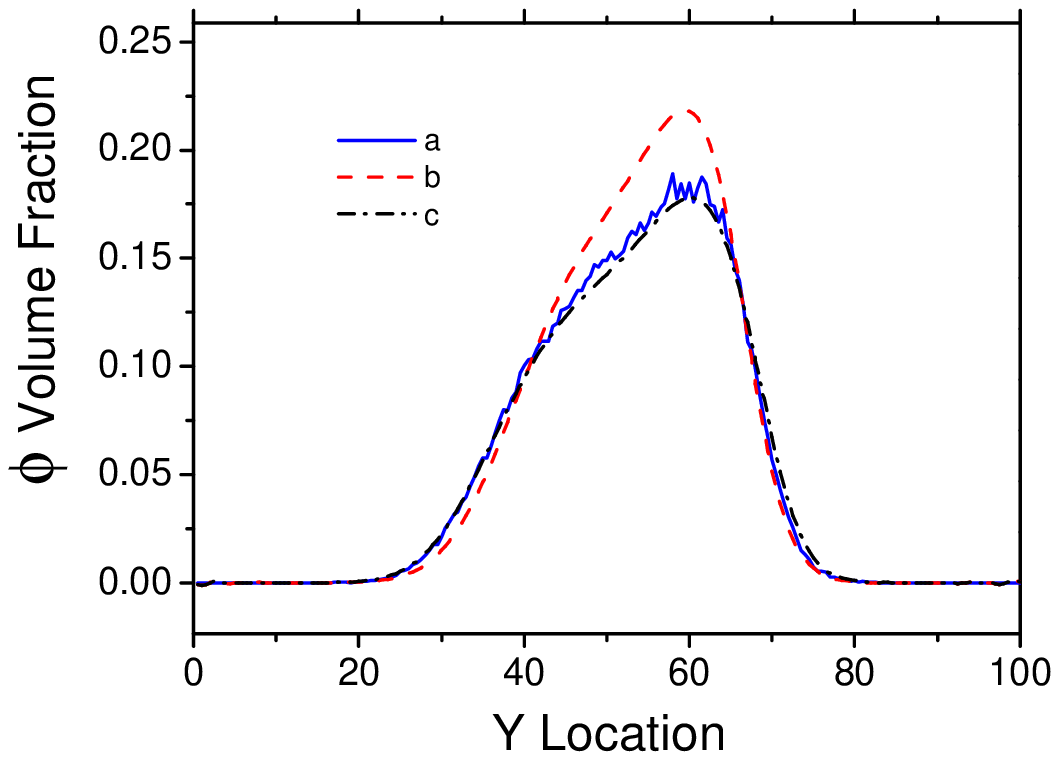}
\end{center}
\caption{Cross sections of the volume fraction $\phi =
L_x^{(0)}L_y^{(0)}p(x_0,y,t)$ in 2D for (a) CPM simulation,  (b) Eq.
(\ref{pottscontinuous3}) and (c) Eq.
({\ref{pottscontinuous4rectangles}}). All cross sections are at the
same position: $x_0 = 57.75$.}\label{figTwoDCross}
\end{figure}

Figures \ref{figTwoD} and \ref{figTwoDCross} demonstrate a very good
agreement between typical CPM simulation and numerical solution of
the continuous model Eq. (\ref{pottscontinuous4rectangles}). Both
simulations were performed on a square domain $0\le x,y \le 100$
over simulation time $t_{end}=400$. Simulation parameters' values
are as follows
 $\triangle r=1$, $L_{T_x}=L_{T_y}=4.4$,
$\lambda_x=\lambda_y=1.5$, $J_{cm}=2$, $\beta=15$, $\mu=0.1$,
$\epsilon= 0.01$ and $N=15$. Chemical field concentration is chosen
in the form of $c(x,y)=0.2[1-e^{-\frac{(x-65)^2+(y-60)^2}{144}}]$
and does not depend on time. Initial cell density is chosen in the
form of $p_0(x,y)=k_0e^{-[\frac{(x-50)^2+(y-50)^2}{100}]^5}$, where
$k_0$ is a constant that normalizes the integral of the cell density
to $N=15$. Numerical solution of the continuous model has been
obtained using pseudo-spectral scheme with $200\times 200$ Fourier
modes.  A large number of CPM simulations (600000) have been run on
a parallel computer cluster to guarantee a representative
statistical ensemble.

Numerics for Eq.   (\ref{pottscontinuous3}) significantly differs
from CPM simulations indicating that excluded volume interactions
are important in the chosen range of values of parameters.

\subsubsection{Application to Vasculogenesis}

\begin{figure}
\begin{center}
\includegraphics[width=1.0\linewidth]{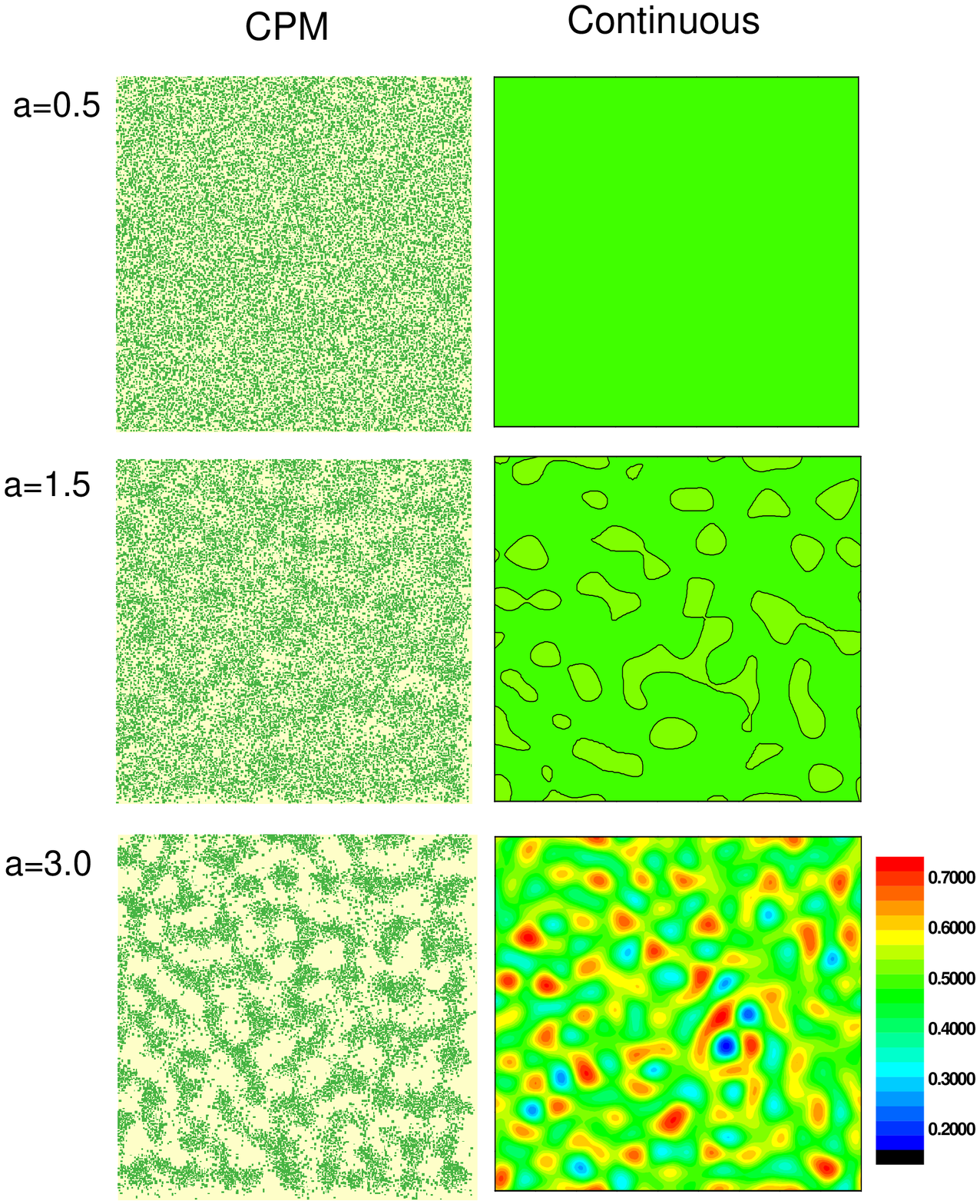}
\caption{Simulation of early vascular network formation with
different chemical production rates $\alpha$. $\triangle r=1$,
$L_{T_x}=L_{T_y}=0.6$, $\lambda_x=\lambda_y=1.5$, $J_{cm}=0.002$,
$\beta=15$, $\mu=-0.1$, $D_c=0.5$, $\gamma=0.014$, $\Delta
t_c=\epsilon^2\Delta t=0.01, \epsilon= 0.1$, $t_{end}=60$. In the
Monte Carlo CPM simulation $ N=15000$ cells were randomly
distributed in a domain $0\le x,y \le 100$ with initial chemical
field at zero. In the continuous model, a uniform initial cell
density distribution with $5\%$ random fluctuation was used. Scale
bar is in units of the volume fraction $\phi$.} \label{figBone}
\end{center}
\end{figure}

To test the model we studied effect of the chemical production rate
on the network formation. Our previous results
\cite{AlberChenLushnikovNewmanPRL2007} were limited to relatively
small chemical production rate $a\le 0.2$ in Eq.
(\ref{ceqmacroscopic1}) because otherwise chemotaxis resulted in
cellular density which was too high for applying mean-field
approximation. Here we use macroscopic Eqs.
(\ref{pottscontinuous4rectangles}), (\ref{ceqmacroscopic1}) and
compare numerical results with the CPM simulations. Figure
\ref{figBone} shows series of simulations with different chemical
production rates $a=0.5, \ a=1.5$ and $a=3.0$. Simulations start
with initially dilute populations of cells moving on a substrate in
a chemotactic field, subject to an excluded volume constraint. Both
CPM and continuous model simulation results indicate that stripe
patterns are obtained for high chemical production rates. Higher
chemical production rate, by strengthening the chemotaxis and cell
aggregation process, eventually leads to higher pattern density with
smaller average distance between two neighboring stripes. Structures
of the resulting networks obtained using discrete and continuous
models are very similar to each other as well as to the one obtained
experimentally for a population of endothelial cells cultured on a
Matrigel film \cite{Serini}.

%\begin{widetext}

\section{Summary and discussion}
We have derived macroscopic continuous Eqs.
(\ref{pottscontinuous4rectangles1Dphi}) and
(\ref{pottscontinuousexcluded2Dphi}) coupled with  Eq.
(\ref{ceqmacroscopic1}) for describing evolution of cellular density
in the chemical field, from microscopic  cellular dynamics. Microscopic cellular model
includes many individual cells moving on a substrate by means of random
fluctuations of their shapes, chemotactic and contact cell-cell interactions. Contrary to classical Keller-Segel model, solutions of
the obtained Eqs. (\ref{pottscontinuous4rectangles1Dphi}),
(\ref{pottscontinuousexcluded2Dphi}) (\ref{ceqmacroscopic1}) do not
collapse in finite time and can be used even when relative volume
occupied by cells $\phi$ is quite large. This makes them much more
biologically relevant then earlier introduced systems. We compared
simulations of macroscopic Eqs. with Monte Carlo simulations of
microscopic cellular dynamics for the CPM and demonstrated a very
good agreement for $\phi\simeq 0.3$. For larger density we expect
transition to glass state \cite{JonesPusey1991}. It was demonstrated
that combination of the CPM and derived continuous model can be
applied to studying network formation in early vasculogenesis. We
are currently working on an important problem  in vasculogenesis of
simulating self-diffusion \cite{JonesPusey1991} of one type of cells
through dense population of other types of cells.

This work was partially supported by NSF grants DMS 0719895 and
IBN-0344647.

\end{document}